\documentclass[sn-mathphys]{sn-jnl}
\jyear{2022}
\theoremstyle{thmstyleone}

\theoremstyle{thmstyletwo}

\theoremstyle{thmstylethree}

\usepackage{multirow,color,bm,bigstrut,booktabs,rotating,hyperref}

\begin{document}

\title[Image Compressed Sensing with Multi-scale Dilated CNN]{Image Compressed Sensing with Multi-scale Dilated Convolutional Neural Network}

\author[1]{\fnm{Zhifeng} \sur{Wang}}\email{zfwang@ccnu.edu.cn}

\author[2]{\fnm{Zhenghui} \sur{Wang}}\email{625612924@qq.com}

\author*[2]{\fnm{Chunyan} \sur{Zeng}}\email{cyzeng@hbut.edu.cn}

\author[2]{\fnm{Yan} \sur{Yu}}\email{1196642246@qq.com}

\author[2]{\fnm{Xiangkui} \sur{Wan}}\email{xkwan@hbut.edu.cn}

\affil[1]{\orgdiv{Department of Digital Media Technology}, \orgname{Central China Normal University}, \city{Wuhan}, \postcode{430079}, \country{China}}

\affil*[2]{\orgdiv{Hubei Key Laboratory for High-efficiency Utilization of Solar Energy and Operation Control of Energy Storage System}, \orgname{Hubei University of Technology}, \city{Wuhan}, \postcode{430068}, \country{China}}

\abstract{Deep Learning (DL) based Compressed Sensing (CS) has been applied for better performance of image reconstruction than traditional CS methods. However, most existing DL methods utilize the block-by-block measurement and each measurement block is restored separately, which introduces harmful blocking effects for reconstruction. Furthermore, the neuronal receptive fields of those methods are designed to be the same size in each layer, which can only collect single-scale spatial information and has a negative impact on the reconstruction process. This paper proposes a novel framework named Multi-scale Dilated Convolution Neural Network (MsDCNN) for CS measurement and reconstruction. During the measurement period, we directly obtain all measurements from a trained measurement network, which employs fully convolutional structures and is jointly trained with the reconstruction network from the input image. It needn't be cut into blocks, which effectively avoids the block effect. During the reconstruction period, we propose the Multi-scale Feature Extraction (MFE) architecture to imitate the human visual system to capture multi-scale features from the same feature map, which enhances the image feature extraction ability of the framework and improves the performance of image reconstruction. In the MFE, there are multiple parallel convolution channels to obtain multi-scale feature information. Then the multi-scale features information is fused and the original image is reconstructed with high quality. Our experimental results show that the proposed method performs favorably against the state-of-the-art methods in terms of PSNR and SSIM.}


\maketitle

\section{Introduction}\label{sec1}

With the continuously increasing of modern data information, it brings a great challenge to signal acquisition, storage, and transmission technologies \cite{zeng1_i,zeng7_i,zeng8_i}. Nyquist sampling theorem does not meet the needs of practical application \cite{wang2_i,wang4_i,wang6_i}. In recent years, the theory of compressed sensing (CS) proposed by Candes \cite{r1} shows that if the signal is sparse or compressible and the measurement matrix satisfies the Restricted Isometry Property (RIP) condition, the original signal can be accurately restored from less than that of the Nyquist sampling theorem, and it saves a lot of memory for data sampling, transmission, and storage \cite{zeng7_i,wang4_i}.

For traditional compressed sensing algorithms, there are two main categories of CS reconstruction methods: convex relaxation methods \cite{r2,r3} and greedy matching pursuit methods \cite{r4,r8_co}. Convex relaxation methods, which mainly include Interior Point Method (IPM) \cite{r6}, Gradient Projection for Sparse Reconstruction (GPSR) \cite{r7}, and Iterative Soft Thresholding Algorithm (ISTA) \cite{r8}, can solve the optimization problem of compressed sensing based on gradient descent \cite{r2}. IPM uses the preconditioned conjugate gradients algorithm to compute the search direction. It can efficiently settle large dense problems, which arise in sparse signal recovery with orthogonal transforms, by exploiting fast algorithms for those transforms. GPSR is based on the gradient descent method. It introduces hidden variables which transform the non-differentiable optimization function into a differentiable unconstrained convex function to reconstruct the original signal. ISTA uses the contraction thresholding function to solve the sub-optimization problem instead of the optimization problem and sets a fixed threshold to select the support set of meeting the conditions. But the calculation of these algorithms is very complicated and the calculation speed is slow.

To accelerate the convergence speed of the algorithms, researchers propose some greedy matching pursuit methods. Representative algorithms are the Orthogonal Matching Pursuit (OMP) algorithm \cite{r8_omp} and Compressive Sampling Matching Pursuit (CoSaMP) \cite{r8_co}. As these methods converge faster and are easy to implement in practice, researchers continue to study these methods. Needell et al. propose Regularized Orthogonal Matching Pursuit (ROMP) algorithm \cite{r9}. It is faster than OMP \cite{r8_omp}, but the stability becomes worse. Kang et al. propose an adaptive subspace OMP method \cite{r10} which utilizes the prior knowledge of target size and coherence of target distribution to change the structure of subspace adaptively. Furthermore, it takes advantage of OMP \cite{r8_omp}, SP \cite{r10_sp}, and SaMP \cite{r10_sasp} to improve the performance of reconstruction. Davenport describes a variant of the CoSaMP algorithm \cite{r11} which uses the D-RIP (a condition on the CS matrix analogous to restricted isometry property). This method mainly focuses on an orientation around recovering the signal rather than its dictionary coefficients. Zhang et al. combine the CoSaMP \cite{r8_co} and the genetic algorithm (GA) \cite{r11_ga} and propose new signal recovery framework \cite{r12} which has better reconstruction quality and effectively avoids premature convergence. Although the performance of traditional algorithms is improved in reconstruction speed and quality, these methods have high computational complexity, and the reconstruction accuracy is limited.

Recently, methods based on deep learning (DL) have been widely applied in multi-media tasks ranging from image classification \cite{r13}, object detection \cite{r14}, recognition \cite{r15}, image super-resolution \cite{r16_sda}, CS image reconstruction \cite{zeng1_i}, speaker recognition \cite{wang5_s,wang7_s,wang8_s,zeng2_s}, to digital forensics \cite{wang1_ad,wang3_ad,wang9_ad,zeng3_ad,zeng4_ad,zeng5_ad,zeng6_ad}. Stacked Denoising Autoencoders (SDA) \cite{r16_sda} is the first deep learning technology applied in the CS field. Mousavi et al. \cite{r17} apply SDA to solve the CS recovery problem, which captures statistical dependencies between different elements of image signals to improve image reconstruction quality. Since convolutional neural network (CNN) \cite{r17_cnn} has achieved great results in image processing, CNN is also applied in CS. Kulkarni et al. \cite{r18} are inspired by SRCNN \cite{r19} and propose a non-iterative reconstruction network (ReconNet), which uses CNN to learn the mapping from CS measurement to the original image. In \cite{r20}, Bo et al. propose FompNet based on CNN, which is used as post-processing for fast matching pursuit algorithm. In \cite{r21}, Zhang et al. are inspired by the Iterative Shrinkage-Thresholding Algorithm (ISTA) to propose ISTA-Net which casts ISTA into deep network form for image CS reconstruction. After the deep Residual Network (ResNets) \cite{r22} is proposed, researchers introduce residual learning to the network to improve reconstruction quality. Yao et al. propose a deep residual reconstruction network (DR2-Net) \cite{r23} which increases network depth based on ReconNet to further improve image reconstruction quality. All the above methods utilize the block-by-block measurement, and each measurement block is restored separately, which ignores the association between image blocks. Therefore, these may produce serious block effects, especially at low measurement rates (e.g., $MR=0.01$). Shi et al. \cite{r23_CS} propose an end-to-end framework dubbed CSNet, which does not directly block the input image during measurement, but uses convolution to obtain information about each image block. CSNet significantly improves image reconstruction quality and achieves fast running speed. In the reconstruction part, these methods only utilize standard CNN whose neuronal receptive fields are designed to the same size in each layer, which is inconsistent with the actual observation of the human visual system, hence hindering the representational ability of CNN.

Compared to the previous convolutional network, the multi-scale network can extract richer feature information and improve the representational ability of CNN. In \cite{r24}, Prabhu et al. propose a multi-scale convolutional network termed U-Finger for fingerprint image denoising and inpainting. U-Finger obtains three different scales by downsampling and upsampling and merges each scale features information that effectively improves image denoising ability. In \cite{r25}, Dong et al. propose a second-order multi-scale super-resolution network that concatenates the output of each RACB module to obtain a multi-scale group and concatenates the output of each group to obtain second-order multi-scale features information. The SMSR network achieves super-resolution reconstruction with high quality, even when dealing with remote sensing image which has highly complex spatial distribution. In \cite{r26} Lian et al. propose multi-scale residual reconstruction network (MSRNet) for CS image reconstruction. Although MSRNet also employs dilated convolution in the reconstruction part, multi-scale feature information is directly fused after extracting feature by dilated convolution with different dilated factors, which does not fully utilize multi-scale feature information.

To solve the above problems, we propose MsDCNN to learn the end-to-end mapping between the original images and the reconstructed images for CS image reconstruction in this paper. Firstly, by completely measuring the original image, we apply a fully convolutional network instead of a traditional CS matrix. The fully convolutional network directly measures the complete image and it doesn't need to be cut into blocks, which effectively uses the image structure information. In the reconstruction period, we design the multi-scale feature extraction (MFE) network architecture which consists of multiple parallel convolutional channels to obtain multi-scale feature information. In each convolutional channel, we apply the dilated convolution with different dilation factors to obtain different receptive fields. Convolutional kernels of different receptive fields can extract different scale feature information after the convolutional operation. The MFE module can not only obtain multi-scale features which provide rich information for subsequent image reconstruction and improve the performance of image reconstruction but also avoid the increase of parameters. The contributions of our research work are mainly in three aspects:
\begin{itemize}
	\item[(1).] We propose a novel multi-scale dilated convolutional neural network for high quality image compressed sensing. The MsDCNN combines and jointly trains the measurement and reconstruction modules to learn the end-to-end mapping between the original image and the reconstructed image, which outperforms many other state-of-the-art methods in the quality of reconstruction.
	\item [(2).] During the measurement period, we train a fully convolutional measurement network to obtain all measurements from the complete input image. Therefore, these adjacent measurement data are closely related to each other, which is totally different from traditional block-by-block measurements. The measurement method effectively uses the structural information of adjacent data to improve the quality of subsequent image reconstruction and eliminates the block effect.
	\item [(3).] During the reconstruction period, in order to improve the feature extraction ability of the traditional deep CS methods with a fixed size feature map, we propose MFE architecture to imitate the human visual system to capture multi-scale feature information, which consists of multiple parallel dilated convolutional channels. We apply dilated convolution with different dilation factors to increase the receptive fields, which capture multi-scale features in the image. Finally, we fuse multiple feature information to further improve the quality of image reconstruction.
\end{itemize}

\begin{table}[t]
	\caption{A list of mathematical notations in this paper.\label{tab0}}
	\centering
	{\begin{tabular*}{20pc}{@{\extracolsep{\fill}}ll@{}}
			\toprule
			\textbf{Notations} & \textbf{Description} \\
			\midrule
			$M$   & the length of the measurement vector \\
			$N$   & the length of the original signal vector \\
			$MR$  & the measurement rate defined as $MR=M/N$ \\
			$\bm{x}$   & an original signal vector with size of $N{\times}1$ \\
			$\bm{y}$   & a $M{\times}1$ measurement vector \\
			$\bm{\Phi}$ & a $M{\times}N$ CS matrix \\
			$\bm{\Psi}$ & a $N{\times}N$ sparse representation matrix \\
			$\bm{s}$   & a $N{\times}1$ sparse transform coefficients vector \\
			$K$   & the sparsity of $\bm{x}$ \\
			$\delta_K$   & the RIP parameter \\
			$d$   & the dilated factor \\
			$B$   & the size of blocks of the input image  \\
			$\bm{X_i}$ & the $i$-$th$ block vecotr of the original signal \\
			$\bm{Y_i}$ & the $i$-$th$ block vecotr of the measurement \\
			$\bm{\Phi_B}$ & the block measurement matrix  \\
			$\bm{X}$   & the original input image matrix  \\
			$\bm{Y}$   & the measurement matrix for image compression  \\
			$\bm{X_1}$ & the initial reconstruction image matrix  \\
			$\bm{X_j}$ & the $j$-$th$ input image matrix  \\
			$\bm{X^*}$ & the reconstruction image matrix \\
			$\bm{\mathcal{M}}$ & a tensor of multi-scale feature maps \\
			$\bm{w_1, w_2, w_s, w_l}$ & the weight matrix of the covolutional layer \\
			$\bm{b}$   & the bias vector of the covolutional layer \\
			$\bm{\Theta}$ & the parameter set of the MsDCNN \\  
			\botrule
	\end{tabular*}}{}
\end{table}

\section{Related work}\label{sec2}
\subsection{Compressed sensing theory}\label{subsec2.1}
The mathematical model of CS measurement is expressed as follows:
\begin{equation}\bm{y} = \bm{\Phi x}\end{equation}
Generally, the above data reconstruction is an ill-posed inverse problem. However, as long as the signal $\bm{x}$ is sparse or compressible, the CS still can recover the signal from the measurement $\bm{y}$. When the original signal is an image, the matrix needs to be flattened into a vector row by row, and then the measurement is performed. After that, the output vector obtained by the reconstruction algorithm is spliced into a reconstruction image matrix. In the sparse domain, the original signal $\bm{x}$ can be presented as follows:
\begin{equation}\bm{x}=\bm{\Psi {s}}\end{equation}
The task of the above signal reconstruction is essentially the problem of $l_{1}$-norm minimization.
\begin{equation}\min \|\bm{s}\|_{1},\ s.t.\  \bm{y=\Phi \Psi {s}}\end{equation}
With $0 < \delta_K < 1$, $\bm{\Phi}$ satisfies the $K$-$order$ RIP condition:
\begin{equation}(1-\delta_K)\|\bm{x}\|_{2}^{2} \leq\|\bm{\Phi x}\|_{2}^{2} \leq(1+\delta_K)\|\bm{x}\|_{2}^{2}
\end{equation}
Then we can reconstruct the original signal $\bm{x}$ from the CS measurement. 
\subsection{Block-based CS measurement }\label{subsec2.2}
\begin{figure}[!t]
	\centering
	\includegraphics[width=3.2in]{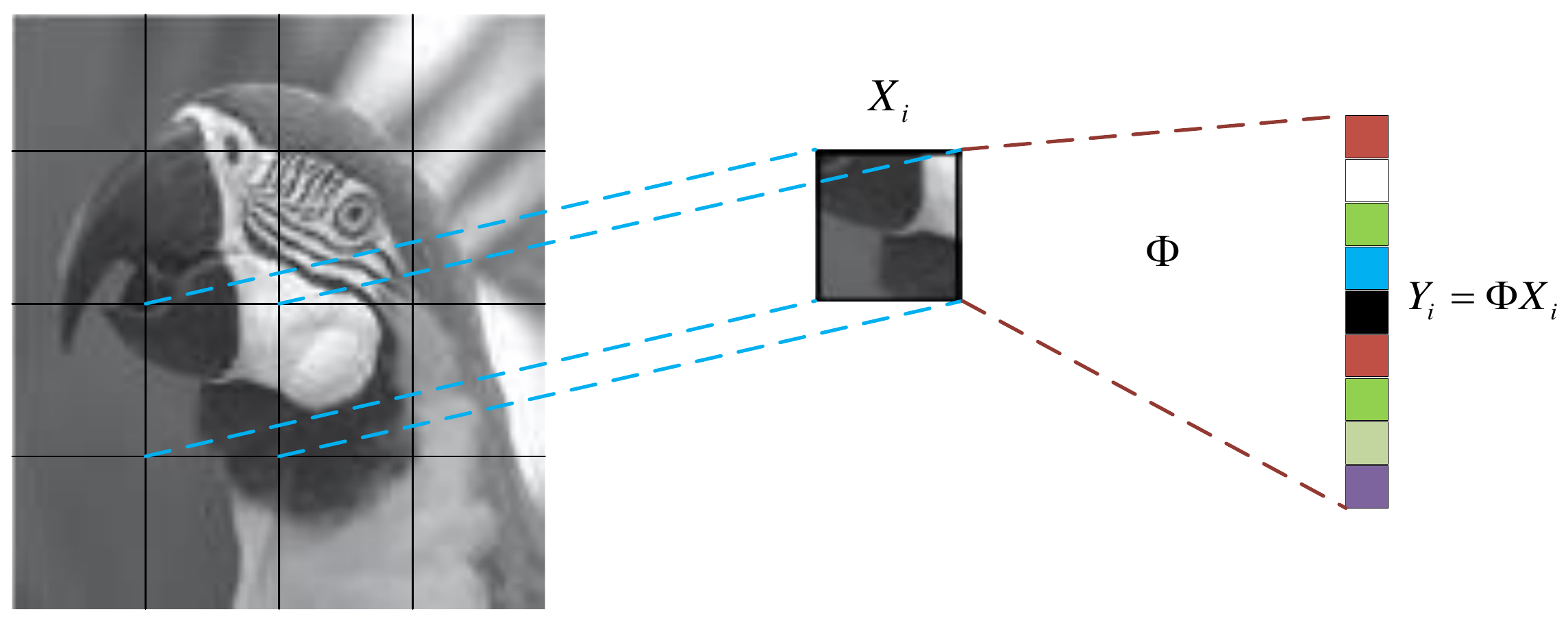}
	\caption{The CS measurement method for each block.}
	\label{fig1}
\end{figure}

Since the measurement can be obtained by $\bm{y=\Phi x}$, the size of the CS matrix increases rapidly with increasing of the input image size. Direct measurement of the whole image requires large storage space and expensive computation. To overcome this problem, the block-based CS method (BCS) \cite{r27} is proposed. The block measurement method of CS is shown in Fig.~\ref{fig1}. In this method, the input image is divided into several non-overlapping blocks and each block uses an independent and smaller CS matrix. The block measurement effectively reduces the dimension of the measurement data and the memory required for the calculation. Although this method effectively reduces computational complexity, it ignores the correlation between adjacent blocks resulting in serious block effects.

\subsection{ Dilated convolution}\label{subsec2.3}
\begin{figure}[!t]
	\centering
	\includegraphics[width=3.2in]{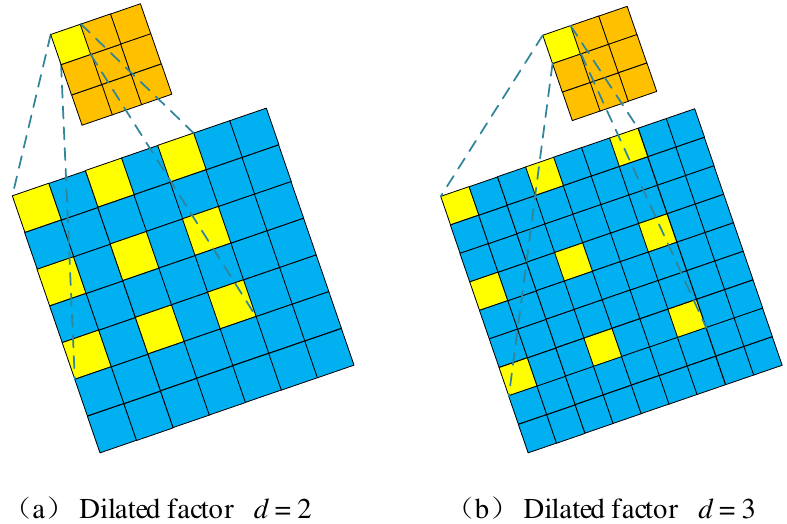}
	\caption{The dilated convolution with different dilated factors.
		\label{fig2}}
\end{figure}

In the process of image reconstruction, it is significant to increase the receptive field. The large receptive field can capture more image information and improve the quality of image reconstruction. In CNN, a large-scale convolutional kernel, pooling layer, and deeper network are generally introduced to increase the receptive field of the network. However, as the size of the convolutional kernel and the number of network layers increase, it will increase the network computational complexity, resulting in a longer image reconstruction time. Although pooling does not increase the computational complexity of the network, it loses a lot of information of the image, resulting in the bad quality of the reconstructed image. Fortunately, the dilated convolution not only increases the receptive field of the network but also maintains computational complexity, which obtains better quality of image reconstruction. For example, using the dilated factor $d=2$ to expand the $3 \times 3$ convolution kernel, which will obtain $(2d+1)\times(2d+1)=5\times5$ convolutional kernel. There are still nine nonzero values in the convolutional kernel, and the values in other positions are zero. The receptive field is changed from the original $3 \times 3$ to $5 \times 5$. Fig.~\ref{fig2} shows dilated convolution with dilated factors $d=2$ and $d=3$.

Although the dilated convolution increases the receptive field without increasing the number of network parameters, it also brings new problems. Since the dilated convolution is a sparse sampling method. When consecutive dilated convolution is used, some pixels are not involved in the calculation, which will lose the continuity and relevance of the data information resulting in the gridding effect. 

\subsection{Motivations}\label{subsec1.1}
To solve the above problems, we propose a novel multi-scale dilated convolution neural network (MsDCNN) for image CS measurement and reconstruction, and the motivations of this paper are as follows:
\begin{enumerate}
	\item [(1).]Most of the previous CS methods use a block-based method to measure. Although the block-based method reduces computational complexity, it also brings some new problems. Since each image block is measured independently in the measurement part, correspondingly, we can only reconstruct each image block respectively. Finally, all reconstructed image blocks are stitched together to obtain a complete image, which causes a serious block effect and obtains bad reconstruction quality.
	\item [(2).]In the reconstruction part, Most of the CS methods based on DL use CNN instead of Deep Neural Network (DNN), which further improves reconstruction performance. But each convolutional layer uses the convolutional kernel of the same receptive field to extract feature information, which can only collect single-scale spatial information, and the potential of CNN is not fully utilized to reconstruct the image with higher quality.
\end{enumerate}

\section{The proposed method}\label{sec3}
\begin{sidewaysfigure}
	\centering
	\includegraphics[width=7.1in]{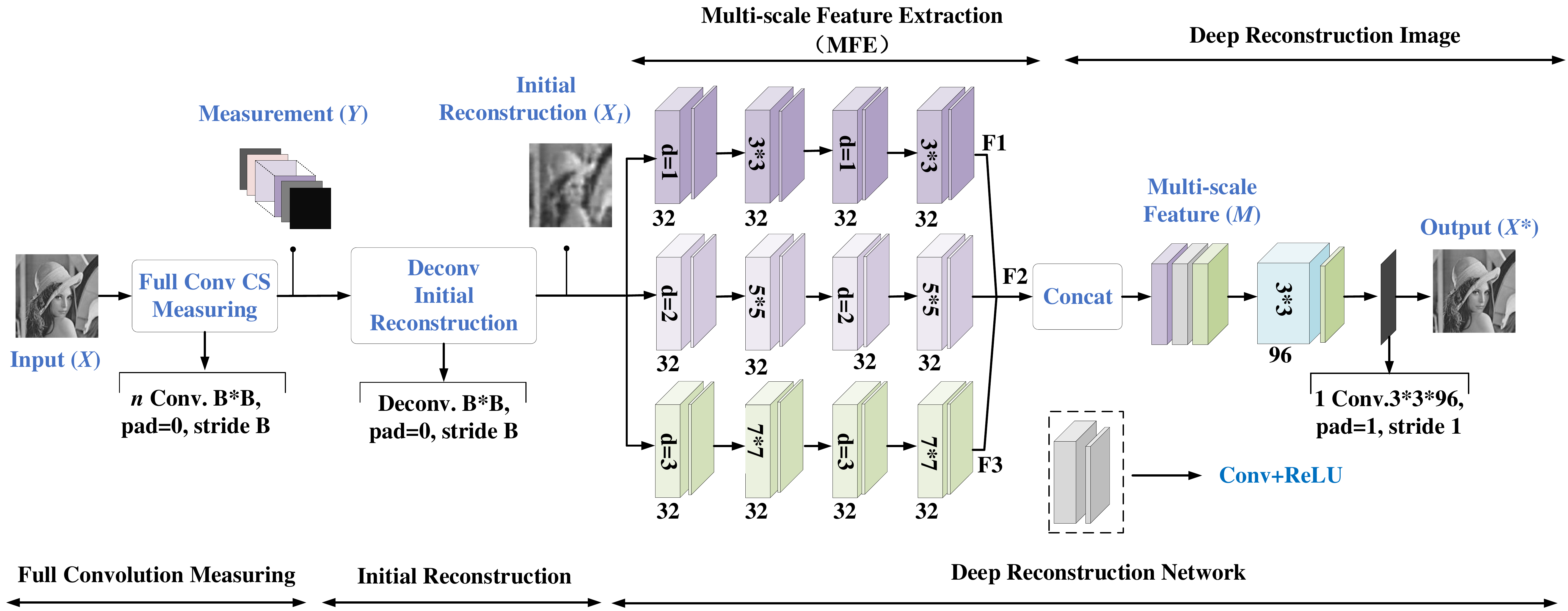}
	\caption{The network architecture of our multi-scale dilated convolutional neural network (take the three parallel channel network as an example).}
	\label{fig3}
\end{sidewaysfigure}

In this section, we propose the MsDCNN to measure and reconstruct the images. As shown in Fig.~\ref{fig3}, the MsDCNN consists of two components, which are the full convolution measurement network and the reconstruction network. The reconstruction network also includes initial reconstruction and deep reconstruction. Since the basic operations of the measurement and reconstruction networks are convolution and deconvolution, which can directly process the input and output image matrix, there is no need to take a matrix-vector conversion in both measurement and reconstruction processes. Then we will describe the details of the network.

\subsection{Network architecture}\label{subsec3.1}

\subsubsection{Full convolution measurement instead of the CS matrix}\label{subsubsec3.1.1}
\begin{figure}[!t]
	\centering
	\includegraphics[width=3.2in]{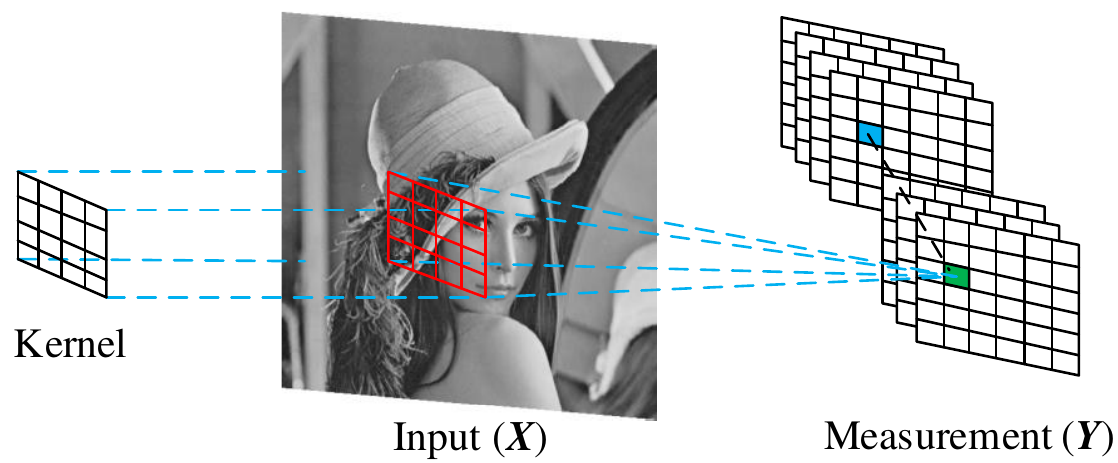}
	\caption{The fully convolutional measurement.}
	\label{fig4}
\end{figure}

The existing CS methods based on deep learning usually adopt the block-by-block measurement method. In the CS measurement, the images are divided into small blocks of $B\times B$, and each small block is measured respectively by the compressed sample expression $\bm{Y_i}=\bm{\Phi_{B}} \times \bm{X_i}$. Here $\bm{\Phi_{B}}$ is usually an artificially designed Gaussian random measurement matrix. The size of the input image block must be fixed which limits practical application, and it will correspondingly produce block effects in the final reconstruction result. In order to overcome this shortcoming, the MsDCNN uses a fully convolutional network to obtain measurement value from the input image, as shown in Fig.~\ref{fig4}.

\begin{equation}\bm{Y}=H_{Cl}(\bm{X})=\bm{w_1} \otimes \bm{X}\end{equation}
Where $H_{Cl}(\cdot)$ denotes convolutional measurement, $\otimes$ denotes convolutional operation, and $\bm{w_1}$ is weights of measurement. Since in the measurement, each convolutional kernel outputs one measurement value. For the measurement rate $M / N$, the CS matrix $\bm{\Phi_B}$ has $n=(M / N) B^{2}$ rows which will obtain $n$ measuring points. We set the number of kernels with $n$ in the measurement layer. In addition, there are no biases in each convolutional kernel and the fully convolutional CS measurement layer has no activation function. The fully convolutional neural network replaces the artificial measurement matrix to adaptively learn the structure information of the input image. And the fully convolutional neural network can adapt to input images of various sizes. 

We summarize the advantages of using a fully convolutional layer for measurement as follows:
\begin{enumerate}
	\item[(1).] It makes full use of the connection between adjacent data and eliminates the block effect caused by block-by-block measurement.
	\item[(2).] It can process images of any size, which breaks the limitation that the fully connected layer can only measure fixed-size images.
\end{enumerate}

\subsubsection{Initial reconstruction with deconvolutional network}\label{subsubsec3.1.2}
After the full convolution measurement, the resolution of the image becomes lower and the size of the images is also compressed. To accurately reconstruct the original image, we first enhance the dimension to the original image size by deconvolution and generate the initial reconstructed image.
\begin{equation}\bm{X_1}=H_{D_e}(\bm{Y})=\bm{w_2} \otimes \bm{y} + \bm{b}
\end{equation}
Where $H_{D_e}(\cdot)$ denotes initial reconstruction, $\bm{w_2}$ and $\bm{b}$ denote the weights and biases respectively of initial reconstruction layer. The deconvolution operation firstly adds zeros to the measured image to expand its dimension to the size of the original image, then transposes the convolution kernel in the previous full convolution measurement, and finally convolves the zero-filled image. Deconvolution is equivalent to the inverse process of convolutional, and deconvolution can be used as up-sampling to improve the dimension to the original image size, which prepares for subsequent deep reconstruction. Although the upsampling method of deconvolution cannot accurately restore the value of the original image, deconvolution has a good ability to learn image features from low level to high level. Therefore, deconvolution is applied as an initial reconstruction network as shown in Fig.~\ref{fig3}. 

\subsubsection{Deep reconstruction network}\label{subsubsec3.1.3}
In order to further improve the quality of image reconstruction, we imitate a human visual system using the multi-channel parallel network. Each channel applies convolutional kernels with different receptive fields to extract different scale information. So after getting the initial reconstructed image, we can obtain the multi-scale features information $(F_1,F_2,F_3)$ via the MFE module. In order to obtain richer feature information, multi-scale feature information is fused by the `Concat' operation.
\begin{equation}\bm{\mathcal{M}}=H_{Concat }(F_1,F_2,F_3)\end{equation}
Where $H_ {Concat }(\cdot)$ denotes concatenate operation, which concatenates different scale channel feature maps by each convolutional channel output to obtain multi-scale features information $\bm{\mathcal{M}}$. The number of channel feature maps of $\bm{\mathcal{M}}$ is the sum of $F_1$, $F_2$ and $F_3$. Finally, we get the final reconstructed image through two convolutional layers as shown in Fig.\ref{fig3}.
\begin{equation}
	\bm{X}^{*}=H_{con}(\bm{\mathcal{M}})=\bm{w_l} \otimes ReLU(\bm{w_s \otimes \mathcal{M}})
\end{equation}
Where $H_{con}(\cdot)$ includes two convolution layers. The $\bm{w_l}$ and $\bm{w_s}$ denote weights of two convolution layers, and the ReLU \cite{r28} is used as activation function. The last convolution layer has no activation function. The $\bm{X^*}$ is the final reconstruction image.

\subsection{The MFE module}\label{subsec3.2}
The MFE is composed of multiple parallel convolutional channels. Each convolutional channel employs different dilated factors to obtain dilated convolutional kernels of different sizes which correspond to different receptive fields. To avoid the gridding effect caused by the continuous dilated convolution of the same dilated factor, we alternately set the dilated convolution and the normal convolution of the same receptive field in each convolutional channel. The MFE module can obtain different scale feature information after convolutional operation for each channel. The multi-scale feature information extracted by the MFE module contains structural information and image details, which provides sufficient information for subsequent multi-feature fusion. So textures, edges, and details of the image can be effectively reconstructed.

\begin{figure*}[!t]
	\centering
	\includegraphics[width=4.7in]{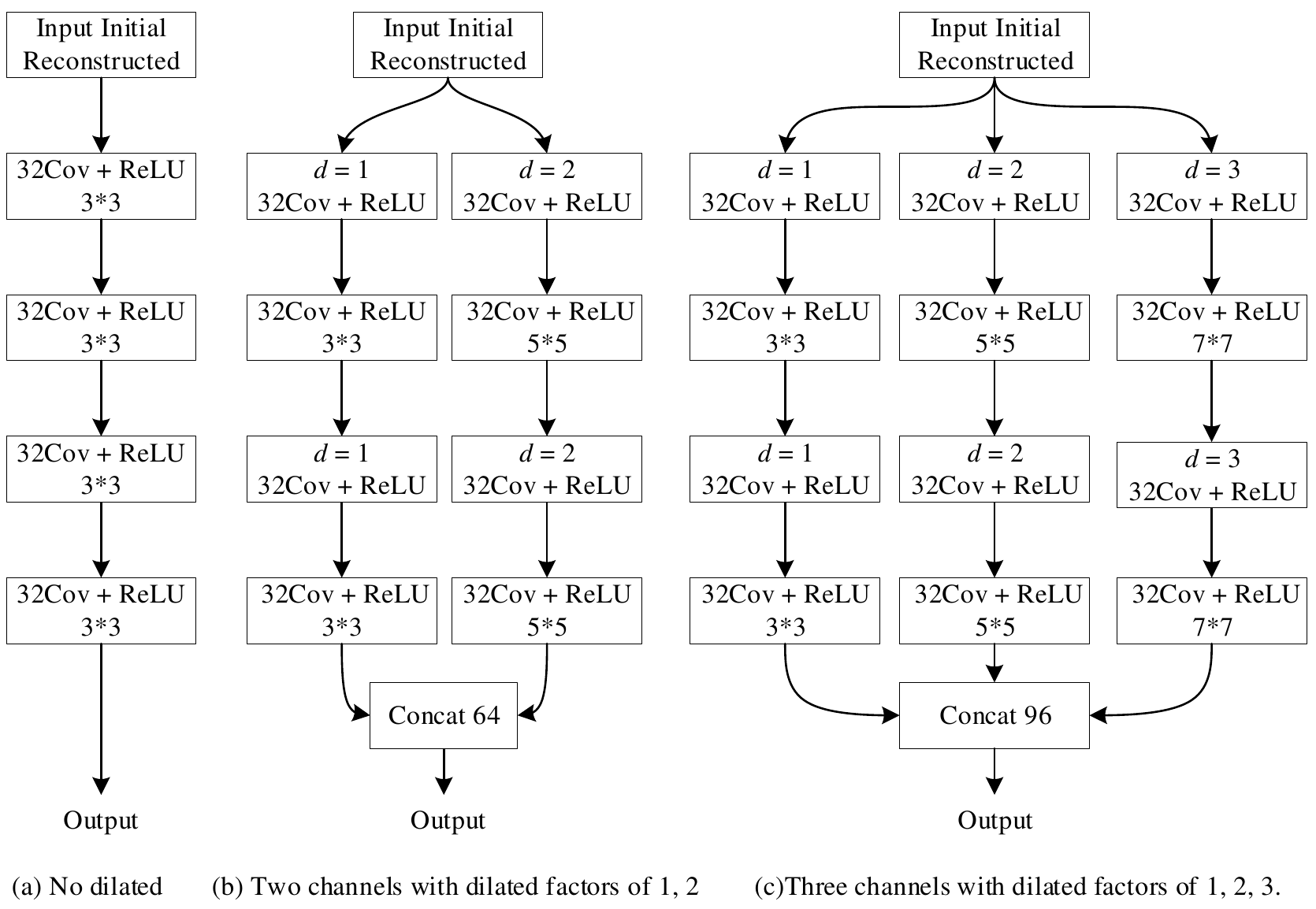}
	\caption{The multi-channel parallel dilated convolution with different dilated factors: ({a}) there is no dilated channel; ({b}) includes two channels with dilated factors of 1 and 2; ({c}) applies three channels with dilated factors of 1, 2, and 3 respectively.}
	\label{fig5}
\end{figure*}

For this module, three situations about channel setting are as follow:
\begin{enumerate}
	\item[(1).] Single channel: This model is similar to the previous network reconstruction model and the size of the convolutional kernel is set to $3\times 3$, which only has a single channel. 
	\item[(2).] Two channels: In the first channel, the dilated factor is 1 liking the single channel. In the second channel, we use dilated convolution with the dilated factor of 2. The receptive field enlarges from $3\times 3$ to $5\times 5$. In order to keep the receptive field consistent and avoid using continuous dilated convolution, we alternately set dilated convolution with the dilated factor of 2 and normal convolution with $5\times 5$.
	\item[(3).] Three channels: On the basis of the two channels, we added the convolutional channel with a dilated factor of 3. The size of convolutional kernel changes from $3\times 3$ to $7\times 7$ which obtains a larger receptive field. Similarly, we alternately set dilated convolution with the dilated factor of 3 and normal convolution with $7\times 7$.
\end{enumerate}
We describe the three different MFE modules and the detailed structure parameters in Fig.~\ref{fig5}.

\subsection{The details of network structure}\label{subsec3.3}
In the measurement part, the size of the convolutional kernel is set to $32\times 32$, The stride is set to 32 for non-overlapping measurement. In the reconstruction part, the kernel size of the deconvolutional layer is $32\times 32$ and the stride also is 32. Each channel of the MFE network has layers, each convolutional layer utilizes 32 convolutional kernels, and the activation function ReLU is used after each convolutional operation. In order to be consistent with the input dimension, we perform a padding operation. After executing MFE, different scale feature information of every channel output is fused to obtain multi-scale features. Fig.~\ref{fig5} shows the structural parameters of MFE in three different cases. In Fig.~\ref{fig5} (a), the MFE is a single channel, which is similar to most other reconstruction networks based on CNN. In Fig.~\ref{fig5} (b), the MFE has two channels, and according to channel dimensions, it merges all outputs to obtain 64 channel feature maps. In Fig.\ref{fig5} (c), the MFE has three channels, which gets 96 channel feature maps. And the final reconstructed image will be obtained through the last two convolutional layers. The size of the convolutional kernel is set to $3\times 3$.

\subsection{Network training}\label{subsec3.4}
Given a training set $\bm{X}$, our goal is to obtain a highly compressed measurement and accurately restore the original input image from measurement. From a practical point of view, it is not feasible to train the measurement network alone. This is because it is difficult to assess how good the quality of the measurement is without the reconstruction error as a reference, so the measurement network and reconstruction network are trained together, which means the CS measurement and reconstruction modules form an end-to-end network structure. The input and the label are images in training the network. Then our network is optimized with a loss function. The goal of training MsDCNN is to minimize the mean square error loss function:

\begin{equation}\min \frac{1}{2I} \sum_{i=1}^{I}\left\|\left(H_{\operatorname{Con}}\left(H_{\text {Concat}}\left(H_{D e}\left(H_{C l}\left(\bm{X_i}, \bm{\Theta}\right)\right)\right)\right)\right)-\bm{X_i}\right\|_{F}^{2}\end{equation}

Where $I$ represents the total number of training samples, $\bm{X_i}$ represents the $i$-$th$ input image,$\bm{\Theta}$ are the parameters that need to be trained, $H_{Cl}(\cdot)$ denotes convolutional measurement operation, $H_{De}(\cdot)$ denotes deconvolution initial reconstruction operation, $H_{Concat}(\cdot)$ denotes 'Concat' operation, which concatenates multiple output feature information, $H_{Con}(\cdot)$ obtains the final output.

\section{Experimental results and discussion}\label{sec4}
In this section, we will evaluate the performance of the proposed method for CS image reconstruction. Firstly, we describe the data set for training and the training details. Then, we verify the effectiveness of the multi-channel and dilated convolution in MFE. Finally, we will compare the proposed method with the state-of-the-art methods with real-world images. \textcolor{blue}{The source codes are released at \url{https://github.com/CCNUZFW/MsDCNN}.}

\subsection{Datasets for training}\label{subsec4.1}
The experiment uses 200 images as a training set and 200 images as a validation set from the BSDS500 database \cite{r29} for network training. We also use data augmentation (rotation or flip) to prepare the training data. Data augmentation (rotate or flip) is used to increase the data needed for network training, to improve the performance of the network model. We use the same test set of the ReconNet and DR2-Net methods as the benchmark.

\subsection{Training details}\label{subsec4.2}
In this subsection, we use the method described in \cite{r30} to initialize the weights. We also use the Adam \cite{r31} optimization method to optimize all parameters of the network. For Adam's other hyper-parameters, the exponential decay rates of the first and second moment estimates are set to 0.9 and 0.999, respectively. The number of epochs is set to 100, the learning rate of the first 50 epochs is 0.001, that of the 51 to 80 epochs is 0.0001 and that of the remaining 20 epochs is 0.00001. Although increasing the number of epochs may improve the performance of the network model, it also increases the training time of the network. Finally, we take the 100th training epoch as the final testing. We implement our model also using MatConvNet \cite{r32} package on MATLAB2016b and train the model on a GPU NVIDIA Quadro M4000.

\subsection{Effectiveness of multi-channel in the MFE module}\label{subsec4.3}

To test the effectiveness of MFE modules under different channel numbers, three MFE modules are set up to test the performance separately. MsDCNN-1 is a single channel network that is similar to the previous simple reconstruction network. MsDCNN-2 indicates MFE has two channels which add dilated convolution with dilated factors of 1 and 2. MsDCNN-3 indicates MFE has three parallel convolutional channels with dilated factors of 1, 2, and 3. 

\begin{table}[!t]
	\centering
	\caption{The average PSNR($dB$) and SSIM with the different number of channels in MFE.\label{tab1}}
	{\begin{tabular*}{20pc}{@{\extracolsep{\fill}}ccccc@{}}
			\toprule
			Image  & \multirow{2}{*}{MR}  & MsDCNN-1  & MsDCNN-2  & MsDCNN-3  \\
			set &  & PSNR/SSIM & PSNR/SSIM & PSNR/SSIM\\
			\midrule
			
			\multirow{3}{*}{Set5} & 0.01 & 23.74/0.6219 
			& 24.05/0.6391 
			& \textbf{24.15/0.6453}
			\\
			& 0.04 
			& 27.89/0.7816 
			& 28.54/0.8127 
			& \textbf{28.58/0.8136}
			\\
			& 0.10 
			& 31.10/0.8709 
			& 31.64/0.8867 
			& \textbf{31.75/0.8892}
			\\
			\midrule
			\multirow{3}{*}{Set14}& 0.01 & 22.47/0.5408 
			& 22.74/0.5568 
			& \textbf{22.79/0.5612}
			\\
			& 0.04 
			& 25.57/0.6758 
			& \textbf{26.09}/0.6982 
			& 26.07/\textbf{0.6989}
			\\
			& 0.10 
			& 27.96/0.7860 
			& 28.36/0.7981 
			& \textbf{28.47/0.8006}
			\\
			\botrule
	\end{tabular*}}{}
\end{table}

To demonstrate that multi-channel is beneficial for improving image reconstruction quality, we evaluate the CS image reconstruction quality with two widely used image quality metrics: PSNR and SSIM, and compare performances at the measurement rates of 0.01, 0.04, and 0.10. Table~\ref{tab1} shows the average PSNR and SSIM of CS image reconstruction for different channels on 5 test images in set5 \cite{r33} and 14 test images in set14 \cite{r34}. It can be seen that PSNR and SSIM for MsDCNN-2 and MsDCNN-3 are higher than the single-channel MsDCNN-1 without dilated convolution. The average PSNR of MsDCNN-2 increases by 0.45$dB$, and the average PSNR of MsDCNN-3 increases by 0.51$dB$ compared with MsDCNN-1. Because multi-channel networks can capture different scale features from the same feature map, the fused multi-scale feature contains richer image information than a single-channel networks. Accordingly, the original image can be reconstructed with high quality. Experimental results also indicate that multi-channel dilated convolution is beneficial to improve image reconstruction quality.

As the dilated factor increases, the receptive field expands correspondingly. However, when the number of parallel convolutional channels exceeds three, the improvement of reconstruction quality is little. In experiments, we design four parallel convolutional channels (The dilated factor of the fourth channel is 4, and the corresponding dilated convolutional kernel is $9\times 9$). The average PSNR of the four channels is only 0.02$dB$ higher than that of the three channels. Although a larger dilated factor (e.g., $d=4$) still can improve the quality of image reconstruction, the improvement effect is not obvious. The reason for this problem is that it will lose more spatial information for an image, as the receptive field continues to expand. And as the number of channels increases, the parameters of the network also increase, which increases the computational complexity of the network. Therefore, we use $d=3$ as the largest dilated factor in MFE.

\subsection{Effectiveness of dilated convolution in the MFE}\label{subsec4.4}
We use dilated convolution with different dilated factors in MFE to increase the receptive field. In order to show that dilated convolution is beneficial, we set different convolutional kernels to experiment in MFE. MsDCNN-2$^{(d)}$ indicates continuous dilated convolution is used in each convolutional layer in MFE. MsDCNN-2$^{(c)}$ uses general convolutional in MFE. MsDCNN-2 alternately sets dilated convolution and general convolutional in each parallel convolutional channel as shown in Fig.~\ref{fig5}(b).

\begin{table}[!t]
	\centering
	\caption{The number of network parameters in MFE.\label{tab2}}
	{\begin{tabular*}{20pc}{@{\extracolsep{\fill}}cccc@{}}
			\toprule
			Model &  MsDCNN-2$^{(d)}$ &  MsDCNN-2$^{(c)}$ & MsDCNN-2\\
			\midrule
			Parameters & \textbf{56k}  & 105k & 88k\\  
			\botrule
	\end{tabular*}}{}
\end{table}

Firstly we calculate the numbers of the parameters of the three methods. It can be seen from Table~\ref{tab2} that the numbers of the parameters of MsDCNN-2$^{(d)}$ is much smaller than that of MsDCNN-2$^{(c)}$, and the number of parameters in MsDCNN-2 is less than that of MsDCNN-2$^{(c)}$ with general convolutional. We also compare the average time cost of the above three methods, as shown in Table~\ref{tab3}. It can be seen, the time cost of MsDCNN-2$^{(d)}$ is the least for reconstructing an image compared with MsDCNN-2$^{(c)}$ and MsDCNN-2, and time cost of MsDCNN-2 is less than MsDCNN-2$^{(c)}$. It indicates that dilated convolution can maintain the number of parameters unchanged and increase the receptive field, thus reducing the computational complexity caused by the extended receptive field.

\begin{table}[!t]
	\centering
	\caption{The average time cost (millisecond) comparisons with MsDCNN-2$^{(d)}$, MsDCNN-2$^{(c)}$ and MsDCNN-2.\label{tab3}}
	{\begin{tabular*}{20pc}{@{\extracolsep{\fill}}ccccc@{}}
			\toprule
			Image	& \multirow{2}{*}{MR} &  MsDCNN-2$^{(d)}$ &  MsDCNN-2$^{(c)}$ & MsRFCNN-2\\
			Set &  & Time Cost & Time Cost & Time Cost\\
			\midrule
			
			\multirow{3}{*}{Set5} & 0.01 & \textbf{40} 
			& 61 
			& 44\\
			& 0.04 
			& \textbf{44} 
			& 53 
			& 46\\
			& 0.10 
			& \textbf{46} 
			& 47 
			& 47
			\\
			\midrule
			\multirow{3}{*}{Set14} & 0.01
			& \textbf{126} 
			& 149
			& 129\\
			& 0.04 
			& \textbf{139} 
			& 150 
			& 147\\
			& 0.10 
			& \textbf{125} 
			& 150 
			& 126\\
			\botrule
	\end{tabular*}}{}
\end{table}

\begin{table}[!t]
	\centering
	\caption{The average PSNR ($dB$) and SSIM comparisons with MsDCNN-2$^{(d)}$, MsDCNN-2$^{(c)}$ and MsDCNN-2.\label{tab4}}
	{\begin{tabular*}{20pc}{@{\extracolsep{\fill}}ccccc@{}}
			\toprule
			Image	& \multirow{2}{*}{MR}	&  MsDCNN-2$^{(d)}$ &  MsDCNN-2$^{(c)}$ &  MsRFCNN-2\\ 
			Set & & PSNR/SSIM & PSNR/SSIM & PSNR/SSIM\\
			\midrule
			
			\multirow{3}{*}{Set5} & 0.01 & 23.87/0.6305 
			& \textbf{24.11/0.6422}
			& 24.05/0.6391\\
			& 0.04 
			& 28.15/0.7928 
			& \textbf{28.55/0.8314}
			& 28.54/0.8127\\
			& 0.10 
			& 31.69/0.8857 
			& \textbf{31.85/0.8912}
			& 31.64/0.8867
			\\
			\midrule
			\multirow{3}{*}{Set14} & 0.01
			& 22.57/0.5486 
			& \textbf{22.78/0.5590}
			& 22.74/0.5568\\
			& 0.04 
			& 25.78/0.6844 
			& \textbf{26.11/0.6995} 
			& 26.09/0.6982\\
			& 0.10 
			& 28.45/0.7978
			& \textbf{28.53/0.8026}
			& 28.36/0.7981\\
			\botrule
	\end{tabular*}}{}
\end{table}

Then, we evaluated the reconstruction quality of MsDCNN-2$^{(d)}$, MsDCNN-2$^{(c)}$ and MsDCNN-2 using the parameters PSNR and SSIM, as shown in Table~\ref{tab4}. It can be seen that MsDCNN-2$^{(d)}$ has the worst reconstruction performance. That's because continuous dilated convolution with the same dilated factor leads to the loss of internal data structure and spatial information, and the reconstruction quality is limited. 

Through the analysis of Table~\ref{tab3} and Table~\ref{tab4}, we can see that the reconstruction quality of MsDCNN-2$^{(c)}$ is slightly higher than MsDCNN-2, but the time cost of MsDCNN-2 is less than MsDCNN-2$^{(c)}$. So we alternately set dilated convolution and general convolution in each parallel convolutional channel, and it not only effectively improves the quality of image reconstruction but also costs almost no more reconstruction time. The MsDCNN-2 used in this paper has compromised reconstruction performance.

\subsection{Effectiveness of the full convolution measurement}\label{subsec4.50}

In order to verify the influence of the measurement methods on the reconstruction quality, we compared the reconstruction algorithms under the three measurement methods of partial-DCT measurement, random Gaussian measurement, and full convolution measurement. For the first two measurement methods, since the size of the input image is different, the input image is divided into multiple $33 \times 33$ small blocks before measurement and then expanded into column vectors by row. After the measurement, a fully connected layer is up-sampled and deformed into a $33 \times 33$ image as the initial reconstructed image, which is in the same way as ReconNet. The following deep reconstruction network structure is the same as MsDCNN. As shown in Table \ref{tab4_0}, DCT-MsDCNN-1 and DCT-MsDCNN-3 refer to single-channel and three-channel reconstruction algorithms under partial-DCT measurement, G-MsDCNN-1 and G-MsDCNN-3 refer to single-channel and three-channel reconstruction algorithms under random Gaussian measurement, and MsDCNN-3 refers to the proposed reconstruction algorithm under the full convolution measurement. Since ReconNet is very similar in structure to DCT-MsDCNN-1 and G-MsDCNN-1, and the number of network layers is the same, it is listed as the baseline under random Gaussian measurement.

In this experiment, the Gray11 dataset is applied as the testing dataset. As shown in Table \ref{tab4_0}, the reconstruction quality under the random Gaussian measurement is significantly better than the reconstruction quality under the partial-DCT measurement, so the random Gaussian measurement is used by most benchmark algorithms. While compared with the Gaussian random measurement, the full convolution measurement can significantly improve the reconstruction quality, and the PSNR is improved by 3.01$dB$ on average at the three sampling rates. There are three main reasons to explain the results of this experiment: Firstly, the measurement network has learned the data characteristics and adjusted the network parameters to adapt to the input image, while the random Gaussian measurement is independent of the input signal. Secondly, when the Gaussian random measurement is independent of the reconstruction network, the end-to-end network framework and training method closely integrate the links between measurement and reconstruction, and the measurement network adaptively promotes the reconstruction. Thirdly, the full convolution measurement does not divide the input image into blocks, which avoids the blocking effect.

The difference between G-MsDCNN-1 and ReconNet is less than 0.1$dB$, because they use the same measurement method and a similar reconstruction network structure. G-MsDCNN-3 is 0.68$dB$ higher than G-MsDCNN-1 on average, especially 1.29$dB$ higher at a measurement rate of 0.1, indicating that multi-channel can indeed improve the reconstruction quality. Therefore, this paper adopts the end-to-end framework of full convolution measurement and multi-channel reconstruction to improve the quality of reconstruction.

\begin{table}[!t]
	\centering
	\caption{The average PSNR ($dB$) for the different measuring methods on the testing dataset of Gray11.\label{tab4_0}}
	{\begin{tabular*}{20pc}{@{\extracolsep{\fill}}cccc@{}}
			\toprule
			Methods & MR=0.10 & MR=0.04 & MR=0.01 \\
			\midrule
			ReconNet & 22.68 & 19.99 & 17.27 \\
			DCT-MsDCNN-1 & 16.09 & 15.55 & 15.12 \\
			DCT-MsDCNN-3 & 16.62 & 15.98 & 15.21 \\
			G-MsDCNN-1 & 22.58 & 19.81 & 17.16 \\
			G-MsDCNN-3 & 23.87 & 20.29 & 17.43 \\
			MsDCNN-3 & \textbf{26.43} & \textbf{23.96} & \textbf{20.22} \\
			\botrule
	\end{tabular*}}{}
\end{table}

\subsection{Comparisons with the state-of-the-art methods}\label{subsec4.5}
\begin{table}[!t]
	\centering
	\caption{The PSNR ($dB$) at different measurement rates for different methods.\label{tab5}}
	{\begin{tabular*}{20pc}{@{\extracolsep{\fill}}ccccc@{}}
			\toprule
			Image Name	& Methods 	& MR=0.10 & MR=0.04   & 
			MR=0.01\\
			\midrule
			
			\multirow{6}{*}{Brbara} 
			& TVAL3 & 21.88 & 18.98 & 11.94\\
			& ReconNet & 21.89 & 20.38 & 18.61\\
			& DR2-Net & 22.69 & 20.70 & 18.65\\
			& MSRNet & 23.04 & 21.01 & 18.60\\
			& MsDCNN-2 & 24.26 & \textbf{23.53} & 21.75 \\
			& MsDCNN-3 & \textbf{24.28} & \textbf{23.53} & \textbf{21.78} \\
			\midrule
			
			\multirow{6}{*}{Boats} 
			& TVAL3 & 23.86 & 19.20 & 11.86\\
			& ReconNet & 24.15 & 21.36 & 18.61\\
			& DR2-Net & 25.58 & 22.11 & 18.67\\
			& MSRNet & 26.32 & 22.58 & 18.65\\
			& MsDCNN-2 & 28.83 & 25.98 & \textbf{21.88} \\
			& MsDCNN-3 & \textbf{28.96} & \textbf{26.05} & \textbf{21.88} \\
			\midrule
			
			\multirow{6}{*}{Flinstones} 
			& TVAL3 & 18.88 & 14.88 & 9.75\\
			& ReconNet & 18.92 & 16.30 & 13.96\\
			& DR2-Net & 21.09 & 16.93 & 14.01\\
			& MSRNet & 21.72 & 17.28 & 13.83\\
			& MsDCNN-2 & 22.91 & 20.07 & 16.57 \\
			& MsDCNN-3 & \textbf{22.98} & \textbf{20.08} & \textbf{16.64} \\
			\midrule
			
			\multirow{6}{*}{Lena} 
			& TVAL3 & 24.16 & 19.46 & 11.87\\
			& ReconNet & 23.83 & 21.28 & 17.87\\
			& DR2-Net & 25.39 & 22.13 & 17.97\\
			& MSRNet & 26.28 & 22.76 & 18.06\\
			& MsDCNN-2 & 28.46 & 25.92 & 22.31 \\
			& MsDCNN-3 & \textbf{28.53} & 25.91 & \textbf{22.44} \\
			\midrule
			
			\multirow{6}{*}{Monarch} 
			& TVAL3 & 21.16 & 16.73 & 11.09\\
			& ReconNet & 21.10 & 18.19 & 15.39\\
			& DR2-Net & 23.10 & 18.93 & 15.33\\
			& MSRNet & 23.98 & 19.26 & 15.41\\
			& MsDCNN-2 & 27.22 & 23.89 & 17.81 \\
			& MsDCNN-3 & \textbf{27.46} & 23.80 & \textbf{18.00} \\
			\midrule 
			
			\multirow{6}{*}{Peppers} 
			& TVAL3 & 22.64 & 18.21 & 11.35\\
			& ReconNet & 22.15 & 19.56 & 16.82\\
			& DR2-Net & 23.73 & 20.32 & 16.90\\
			& MSRNet & 24.91 & 20.90 & 17.10\\
			& MsDCNN-2 & 26.24 & 24.34 & 20.56 \\
			& MsDCNN-3 & \textbf{26.59} & 24.31 & \textbf{20.63} \\
			\midrule
			
			\multirow{6}{*}{Mean PSNR} 
			& TVAL3 & 22.84 & 18.39 & 11.31\\
			& ReconNet & 22.68 & 19.99 & 17.27\\
			& DR2-Net & 24.32 & 20.80 & 17.44\\
			& MSRNet & 25.16 & 21.41 & 17.54\\
			& MsDCNN-2 & 26.32 & 23.96 & 20.15 \\
			& MsDCNN-3 & \textbf{26.43} & \textbf{23.96} & \textbf{20.22} \\
			\botrule
	\end{tabular*}}{}
\end{table}

We compare our method MsDCNN-2 and MsDCNN-3 with traditional algorithms TVAL3\cite{r35} and DL algorithms including ReconNet, DR2-Net, and MSRNet. The number of reconstruction network layers of DL-based algorithms are 6, 12, and 7 respectively, and our reconstruction network has 7 layers.
\begin{figure*}[!h]
	\centering
	\includegraphics[width=4.2in]{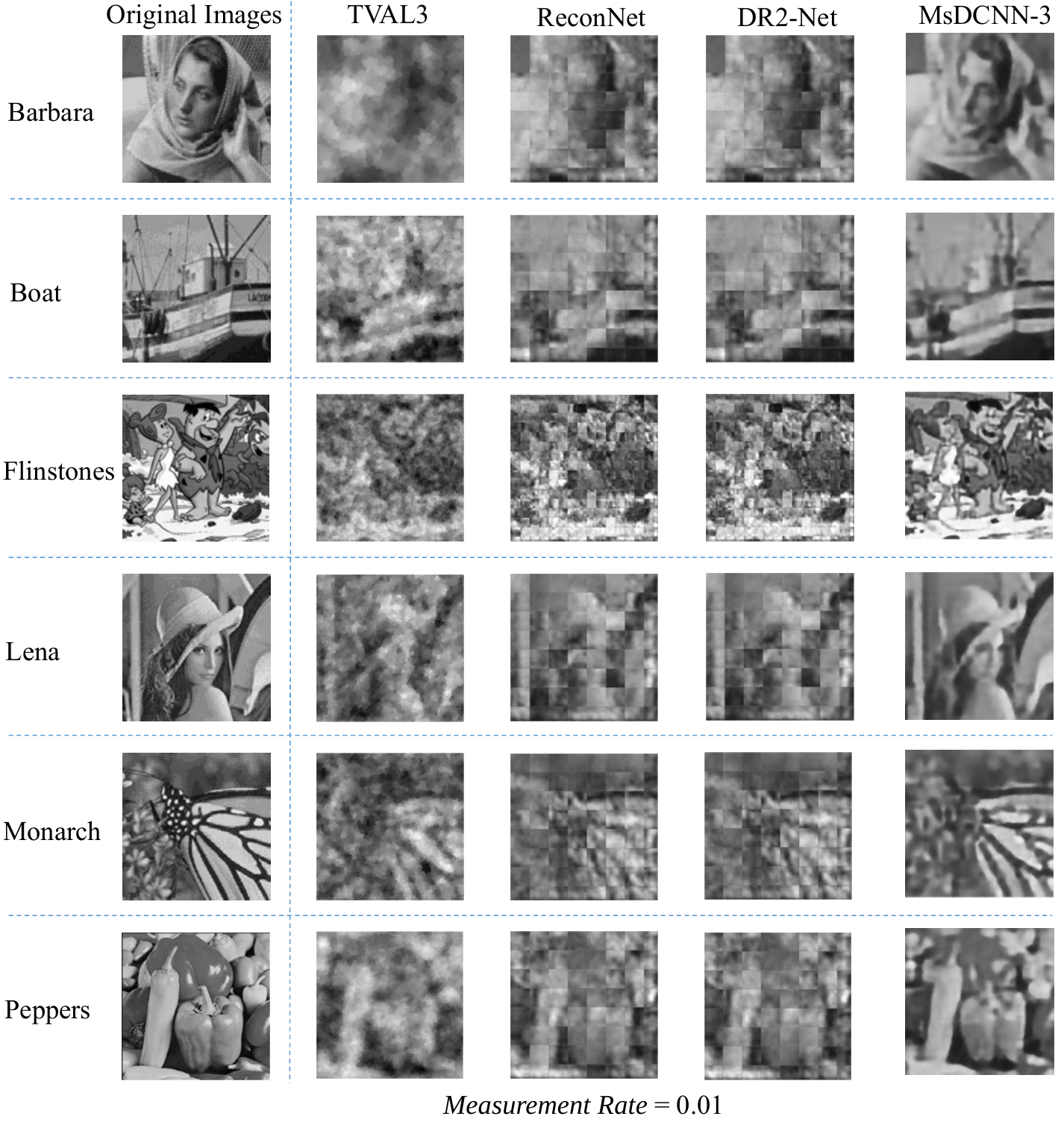}
	\caption{When $MR = 0.01$, the reconstruction images are captured for visual comparison.}
	\label{fig6}
\end{figure*}

Firstly, we use PSNR as the evaluation parameter to quantitatively compare these methods. Table~\ref{tab5} shows the PSNR of different algorithms at different measurement rates. The results of TVAL3\cite{r35} and MSRNet are provided by literature \cite{r26}. As shown in this experiment, the DL based CS algorithms are better than traditional algorithms, and our method has the best performance at all measurement rates. Although our network has more layers than ReconNet, the PSNR of our method is higher than that of ReconNet by about 3.5 $dB$. Moreover, the number of our reconstruction network layers is equal to MSRNet and much less than DR2- Net, but PSNR is obviously improved than MSRNet and DR2- Net. At a low measurement rate of 0.01, the advantage of our method is particularly prominent. At the same time, we also compare MsDCNN-2 and MsDCNN-3 and found that most of the time, the quality of three-channel reconstruction is slightly higher than that of two-channel reconstruction. Experimental results further show that multi-channel parallel expansion convolutional is beneficial to improving image reconstruction quality.

\begin{figure*}[!t]
	\centering
	\includegraphics[width=4.2in]{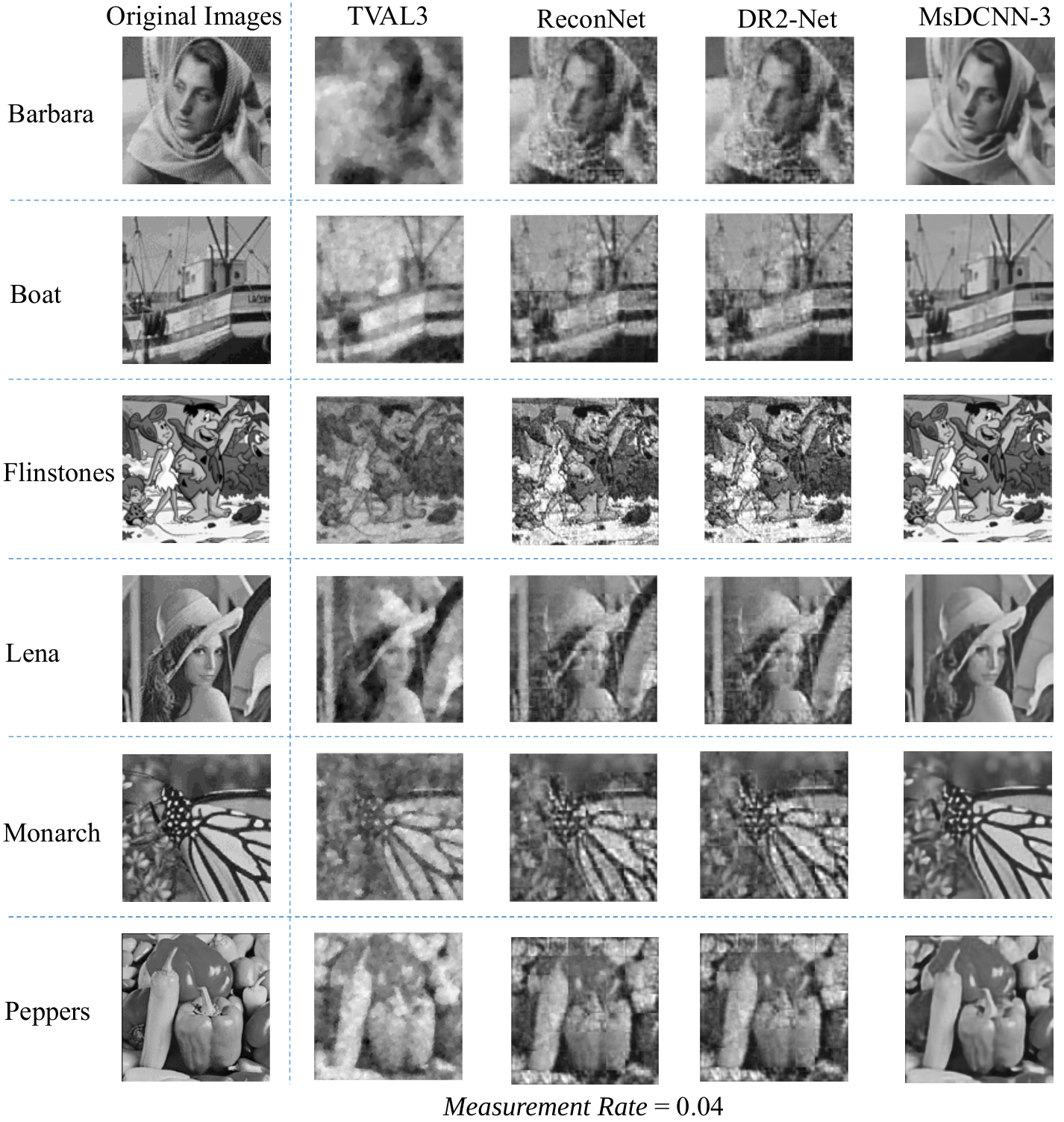}
	\caption{When $MR = 0.04$, the reconstruction images are captured for visual comparison.}
	\label{fig7}
\end{figure*}

Then, We compare the time complexity of these deep learning methods. It can be seen from Table~\ref{tab6} that when reconstructing a single image, the average time cost of our method is higher than that of the state-of-art methods. This is because these methods divide the image into small blocks before inputting the network. The dimension of the image block is much smaller than that of the complete image. Therefore, it is easier to reconstruct the image blocks separately than the complete image. At the same time, we can also see that the time cost of reconstructing an image is still slightly increasing as the number of channels increases. The reason for this problem is that it will bring other additional parameters as we add more parallel convolutional channels. Although our time cost is slightly higher than that of other methods, the reconstruction quality of the image is greatly improved.

\begin{figure*}[!t]
	\centering
	\includegraphics[width=4.2in]{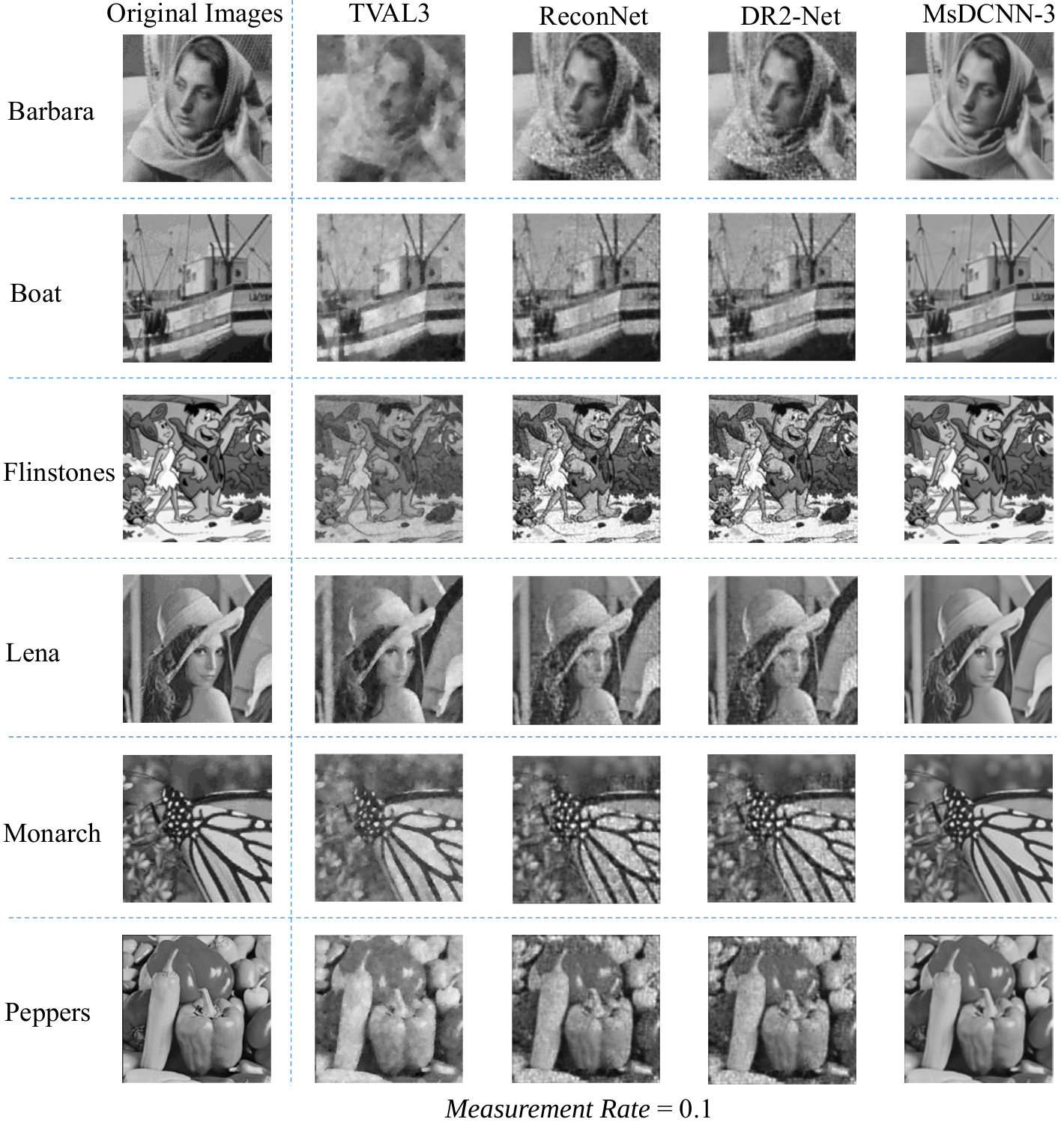}
	\caption{When $MR = 0.1$, the reconstruction images are captured for visual comparison.}
	\label{fig8}
\end{figure*}

\begin{table}[!h]
	\centering
	\caption{The average time cost (millisecond) for reconstruction a single $256\times 256$ image in GPU.\label{tab6}}
	{\begin{tabular*}{20pc}{@{\extracolsep{\fill}}ccccc@{}}
			\toprule
			\multirow{2}{*}{Methods} 	&
			MR=0.01	&   MR=0.04 & 
			MR=0.10 \\
			& Time Cost &  Time Cost & Time Cost \\
			\midrule
			
			ReconNet & \textbf{10.7} & \textbf{10.0} & \textbf{10.1}\\
			DR2-Net & 31.7 & 11.7 & 31.4\\
			MSRNet & 12.1 & 12.4 & 11.7\\
			MsDCNN-1 & 30.0 & 28.3 & 28.1\\
			MsDCNN-2 & 34.0 & 36.4 & 35.3\\
			MsDCNN-3 & 41.2 & 43.1 & 42.2\\
			\botrule
	\end{tabular*}}{}
\end{table}

Finally, we compare our method MsDCNN-3 with TVAL3, ReconNet, and DR2-Net in terms of visual effect, as shown in Fig.~\ref{fig6}, Fig.~\ref{fig7}, and Fig.~\ref{fig8}, in which the CS measurement rates are 0.01, 0.04, and 0.10, respectively. It can be seen from the reconstruction images that our method achieves the best performance in visual effects. Even at a very low measurement rate of 0.01, MsDCNN-3 can effectively eliminate the block effect and retain sharper edges and finer detail.

In Fig.~\ref{fig8}, although the measurement rate is 0.10, there is still a block effect in the region and edges of the person using ReconNet and DR2-Net. The reconstructed image by our method is not distorted in vision compared with the original image. As the measurement rate decreases, we can see that the images reconstructed by ReconNet and DR2-Net become blurry from Fig.~\ref{fig7}. Images have serious block effects in high-frequency areas, and the edges seriously affect the visual effect. When the measurement rate is 0.01 in Fig.~\ref{fig6}, images reconstructed by ReconNet and DR2-Net have severe block effects, and it is even difficult to judge the semantic information. However, the semantic meaning of the image can be clearly inferred from the image reconstructed by our method.

\section{Conclusion}\label{sec5}
In this paper, we propose a multi-scale dilated convolutional neural network for image CS measurement and reconstruction, where fully convolutional is used as CS image measurement and the MFE module serves as multi-scale feature extraction to perform the deep reconstruction. In the measurement part, we use the fully convolutional measurement method instead of the previously block-by-block measurement, in which the measurement matrix can be automatically learned in a trained measurement network. Fully convolutional trained measurement effectively eliminates the block effect caused by the block-by-block measurement and preserves more structure information for subsequent image reconstruction. Specifically, in the reconstruction part, we propose the MFE architecture to imitate the human visual system to capture multi-scale feature information. In the MFE, there are multiple parallel convolutional channels and the dilated convolution are applied to obtain multi-scale receptive fields. In addition, the MFE can capture multi-scale feature information in images to improve the performance of image reconstruction. Experimental results show that the proposed end-to-end CS network achieves significant performance compared with the existing state-of-the-art methods. For future work, we will apply the residual network to obtain the multi-scale feature information of the image, adopt the weighted fusion method to fuse the feature information of different scales to further improve the quality of image reconstruction, and move our experiment implementations to the latest deep learning frameworks. Furthermore, the mathematical proofs of the reconstruction conditions of deep learning-based CS methods are a problem worthy of further study.

\section*{Acknowledgments}\label{sec6}
The research work of this paper were supported by the National Natural Science Foundation of China (No. 62177022, 61901165, 61501199), Collaborative Innovation Center for Informatization and Balanced Development of K-12 Education by MOE and Hubei Province (No. xtzd2021-005), and Self-determined Research Funds of CCNU from the Colleges’ Basic Research and Operation of MOE (No. CCNU22QN013).

\subsubsection*{Data Availability Statement} The datasets analyzed during the current study are available in IEEE Transactions On Pattern Analysis And Machine Intelligence paper ``Contour detection and hierarchical image segmentation'' \cite{r29}.

\vfill\pagebreak


\begin{thebibliography}{999}
\bibitem{r28} Ahn, N., Kang, B., Sohn, K. A. (2018). Fast, accurate, and lightweight super-resolution with cascading residual network. In Proceedings of the European conference on computer vision (ECCV) (pp. 252-268).

\bibitem{r29} Arbelaez, P., Maire, M., Fowlkes, C., Malik, J. (2010). Contour detection and hierarchical image segmentation. IEEE transactions on pattern analysis and machine intelligence, 33(5), 898-916.

\bibitem{r20} Bo, L., Lu, H., Lu, Y., Meng, J., Wang, W. (2017, October). FompNet: Compressive sensing reconstruction with deep learning over wireless fading channels. In 2017 9th International Conference on Wireless Communications and Signal Processing (WCSP) (pp. 1-6). IEEE.

\bibitem{r1} Candès, E. J., Romberg, J., Tao, T. (2006). Robust uncertainty principles: Exact signal reconstruction from highly incomplete frequency information. IEEE Transactions on information theory, 52(2), 489-509. 

\bibitem{r11} Davenport, M. A., Needell, D., Wakin, M. B. (2013). Signal space CoSaMP for sparse recovery with redundant dictionaries. IEEE Transactions on Information Theory, 59(10), 6820-6829.

\bibitem{r31} Deng, Z., Zhu, L., Hu, X., Fu, C. W., Xu, X., Zhang, Q., ... Heng, P. A. (2019). Deep multi-model fusion for single-image dehazing. In Proceedings of the IEEE/CVF international conference on computer vision (pp. 2453-2462).

\bibitem{r19} Dong, C., Loy, C. C., He, K., Tang, X. (2014, September). Learning a deep convolutional network for image super-resolution. In European conference on computer vision (pp. 184-199). Springer, Cham.

\bibitem{r25} Dong, X., Wang, L., Sun, X., Jia, X., Gao, L., Zhang, B. (2020). Remote sensing image super-resolution using second-order multi-scale networks. IEEE Transactions on Geoscience and Remote Sensing, 59(4), 3473-3485.

\bibitem{r13} Fang, L., Wang, C., Li, S., Rabbani, H., Chen, X., Liu, Z. (2019). Attention to lesion: Lesion-aware convolutional neural network for retinal optical coherence tomography image classification. IEEE transactions on medical imaging, 38(8), 1959-1970.

\bibitem{r27} Gan, L. (2007, July). Block compressed sensing of natural images. In 2007 15th International conference on digital signal processing (pp. 403-406). IEEE.

\bibitem{r10_sp} Han, X., Zhao, G., Li, X., Shu, T., Yu, W. (2019). Sparse signal reconstruction via expanded subspace pursuit. Journal of Applied Remote Sensing, 13(4), 046501.

\bibitem{r22} He, K., Zhang, X., Ren, S., Sun, J. (2016). Deep residual learning for image recognition. In Proceedings of the IEEE conference on computer vision and pattern recognition (pp. 770-778).

\bibitem{r10} Kang, L., Huang, J. J., Huang, J. X. (2018, August). Adaptive subspace OMP for infrared small target image. In 2018 14th IEEE International Conference on Signal Processing (ICSP) (pp. 445-449). IEEE.

\bibitem{r17_cnn} Kattenborn, T., Leitloff, J., Schiefer, F., Hinz, S. (2021). Review on Convolutional Neural Networks (CNN) in vegetation remote sensing. ISPRS Journal of Photogrammetry and Remote Sensing, 173, 24-49.

\bibitem{r11_ga} Katoch, S., Chauhan, S. S., Kumar, V. (2021). A review on genetic algorithm: past, present, and future. Multimedia Tools and Applications, 80(5), 8091-8126.

\bibitem{r18} Kulkarni, K., Lohit, S., Turaga, P., Kerviche, R., Ashok, A. (2016). Reconnet: Non-iterative reconstruction of images from compressively sensed measurements. In Proceedings of the IEEE Conference on Computer Vision and Pattern Recognition (pp. 449-458).

\bibitem{r34} Lai, W. S., Huang, J. B., Ahuja, N., Yang, M. H. (2018). Fast and accurate image super-resolution with deep laplacian pyramid networks. IEEE transactions on pattern analysis and machine intelligence, 41(11), 2599-2613.

\bibitem{r3} Li, C., Liu, X., Yu, K., Wang, X., Zhang, F. (2020). Debiasing of seismic reflectivity inversion using basis pursuit de-noising algorithm. Journal of Applied Geophysics, 177, 104028.

\bibitem{r35} Li, C., Yin, W., Jiang, H., Zhang, Y. (2013). An efficient augmented Lagrangian method with applications to total variation minimization. Computational Optimization and Applications, 56(3), 507-530.	

\bibitem{r33} Li, J., Fang, F., Mei, K., Zhang, G. (2018). Multi-scale residual network for image super-resolution. In Proceedings of the European conference on computer vision (ECCV) (pp. 517-532).

\bibitem{r8} Li, W., Niu, M., Zhang, Y., Huang, Y., Yang, J. (2020). Forward-looking scanning radar superresolution imaging based on second-order accelerated iterative shrinkage-thresholding algorithm. IEEE Journal of Selected Topics in Applied Earth Observations and Remote Sensing, 13, 620-631.

\bibitem{r14} Li, Z., Peng, C., Yu, G., Zhang, X., Deng, Y., Sun, J. (2018). Detnet: Design backbone for object detection. In Proceedings of the European conference on computer vision (ECCV) (pp. 334-350).

\bibitem{r26} Lian, Q., Fu, L., Chen, S., Shi, B. (2019). A compressed sensing algorithm based on multi-scale residual reconstruction network. Acta Autom, 45(11), 2082-2091.

\bibitem{r6} Lin, T., Ma, S., Ye, Y., Zhang, S. (2021). An ADMM-based interior-point method for large-scale linear programming. Optimization Methods and Software, 36(2-3), 389-424.

\bibitem{r7} Liu, J. K., Du, X. L. (2018). A gradient projection method for the sparse signal reconstruction in compressive sensing. Applicable Analysis, 97(12), 2122-2131.

\bibitem{r17} Mousavi, A., Patel, A. B., Baraniuk, R. G. (2015, September). A deep learning approach to structured signal recovery. In 2015 53rd annual allerton conference on communication, control, and computing (Allerton) (pp. 1336-1343). IEEE.

\bibitem{r15} Mujahid, A., Awan, M. J., Yasin, A., Mohammed, M. A., Damaševičius, R., Maskeliūnas, R., Abdulkareem, K. H. (2021). Real-time hand gesture recognition based on deep learning YOLOv3 model. Applied Sciences, 11(9), 4164.

\bibitem{r9} Needell, D., Vershynin, R. (2009). Uniform uncertainty principle and signal recovery via regularized orthogonal matching pursuit. Foundations of computational mathematics, 9(3), 317-334.

\bibitem{r24} Prabhu, R., Yu, X., Wang, Z., Liu, D., Jiang, A. A. (2019). U-finger: Multi-scale dilated convolutional network for fingerprint image denoising and inpainting. In Inpainting and Denoising Challenges (pp. 45-50). Springer, Cham.

\bibitem{r2} Saha, T., Srivastava, S., Khare, S., Stanimirović, P. S., Petković, M. D. (2019). An improved algorithm for basis pursuit problem and its applications. Applied Mathematics and Computation, 355, 385-398.

\bibitem{r8_omp} Schnass, K. (2018). Average performance of orthogonal matching pursuit (OMP) for sparse approximation. IEEE Signal Processing Letters, 25(12), 1865-1869.

\bibitem{r23_CS} Shi, W., Jiang, F., Liu, S., Zhao, D. (2019). Image compressed sensing using convolutional neural network. IEEE Transactions on Image Processing, 29, 375-388.

\bibitem{r8_co} Tirer, T., Giryes, R. (2020). Generalizing CoSaMP to signals from a union of low dimensional linear subspaces. Applied and Computational Harmonic Analysis, 49(1), 99-122.

\bibitem{wang1_ad} Wang, Z., Yang, Y., Zeng, C., Kong, S., Feng, S., Zhao, N. (2022). Shallow and deep feature fusion for digital audio tampering detection. EURASIP Journal on Advances in Signal Processing, 2022(1), 1-20.

\bibitem{wang2_i} Wang, Z., Zuo, C., Zeng, C. (2021). SAE based unified double JPEG compression detection system for Web image forensics. International Journal of Web Information Systems. 17(2), 84-98.

\bibitem{wang3_ad} Wang, Z. F., Wang, J., Zeng, C. Y., Min, Q. S., Tian, Y., Zuo, M. Z. (2018, July). Digital audio tampering detection based on ENF consistency. In 2018 International Conference on Wavelet Analysis and Pattern Recognition (ICWAPR) (pp. 209-214). IEEE.

\bibitem{wang4_i} Wang, Z. F., Zhu, L., Min, Q. S., Zeng, C. Y. (2017, July). Double compression detection based on feature fusion. In 2017 International Conference on Machine Learning and Cybernetics (ICMLC) (Vol. 2, pp. 379-384). IEEE.

\bibitem{wang5_s} Wang, Z., Duan, S., Zeng, C., Yu, X., Yang, Y., Wu, H. (2020, November). Robust Speaker Identification of IoT based on Stacked Sparse Denoising Auto-encoders. In 2020 International Conferences on Internet of Things (iThings) (pp. 252-257). IEEE.

\bibitem{wang6_i} Wang, Z., Liu, Q., Yao, H., Chen, J. (2015, October). Virtual chime-bells experimental system based on multi-modal fusion. In 2015 International Conference of Educational Innovation through Technology (EITT) (pp. 64-67). IEEE.

\bibitem{wang7_s} Wang, Z., Duan, S., Zeng, C., Yu, X., Yang, Y., Wu, H. (2020, November). Robust Speaker Identification of IoT based on Stacked Sparse Denoising Auto-encoders. In 2020 International Conferences on Internet of Things (iThings) (pp. 252-257). IEEE.

\bibitem{wang8_s} Wang, Z., Zeng, C., Duan, S., Ouyang, H., Xu, H. (2020, August). Robust Speaker Recognition Based on Stacked Auto-encoders. In International Conference on Network-Based Information Systems (pp. 390-399). Springer, Cham.

\bibitem{wang9_ad} Wang, Z., Liu, Q., Chen, J., Yao, H. (2015, October). Recording source identification using device universal background model. In 2015 International Conference of Educational Innovation through Technology (EITT) (pp. 19-23). IEEE.

\bibitem{r23} Yao, H., Dai, F., Zhang, S., Zhang, Y., Tian, Q., Xu, C. (2019). Dr2-net: Deep residual reconstruction network for image compressive sensing. Neurocomputing, 359, 483-493.

\bibitem{r10_sasp} Yao, S., Guan, Q., Wang, S., Xie, X. (2018). Fast sparsity adaptive matching pursuit algorithm for large-scale image reconstruction. EURASIP journal on wireless communications and networking, 2018(1), 1-8.

\bibitem{r4} Zarei, A., Asl, B. M. (2021). Automatic seizure detection using orthogonal matching pursuit, discrete wavelet transform, and entropy based features of EEG signals. Computers in Biology and Medicine, 131, 104250.

\bibitem{r16_sda} Zeng, C., Ye, J., Wang, Z., Zhao, N., Wu, M. (2022). Cascade neural network-based joint sampling and reconstruction for image compressed sensing. Signal, Image and Video Processing, 16(1), 47-54.

\bibitem{zeng1_i} Zeng, C., Yan, K., Wang, Z., Yu, Y., Xia, S., Zhao, N. (2022). Abs-CAM: a gradient optimization interpretable approach for explanation of convolutional neural networks. Signal, Image and Video Processing, 1-8.

\bibitem{zeng2_s} Zeng, C. Y., Ma, C. F., Wang, Z. F., Ye, J. X. (2018, July). Stacked autoencoder networks based speaker recognition. In 2018 International Conference on Machine Learning and Cybernetics (ICMLC) (Vol. 1, pp. 294-299). IEEE.

\bibitem{zeng3_ad} Zeng, C., Zhu, D., Wang, Z., Wu, M., Xiong, W., Zhao, N. (2021). Spatial and temporal learning representation for end-to-end recording device identification. EURASIP Journal on Advances in Signal Processing, 2021(1), 1-19.

\bibitem{zeng4_ad} Zeng, C., Yang, Y., Wang, Z., Kong, S., Feng, S. (2022). Audio Tampering Forensics Based on Representation Learning of ENF Phase Sequence. International Journal of Digital Crime and Forensics (IJDCF), 14(1), 1-19.

\bibitem{zeng5_ad} Zeng, C., Zhu, D., Wang, Z., Wang, Z., Zhao, N., He, L. (2020). An end-to-end deep source recording device identification system for web media forensics. International Journal of Web Information Systems, 16(4), 413-425. 

\bibitem{zeng6_ad} Zeng, C., Zhu, D., Wang, Z., Yang, Y. (2020, August). Deep and shallow feature fusion and recognition of recording devices based on attention mechanism. In International Conference on Intelligent Networking and Collaborative Systems (pp. 372-381). Springer, Cham.

\bibitem{zeng7_i} Zeng, C., Wang, Z., Wang, Z., Yan, K., Yu, Y. (2021, September). Image Compressed Sensing and Reconstruction of Multi-Scale Residual Network Combined with Channel Attention Mechanism. In Journal of Physics: Conference Series (Vol. 2010, No. 1). IOP Publishing.

\bibitem{zeng8_i} Zeng, C., Wang, Z., Wang, Z. (2020, November). Image Reconstruction of IoT based on Parallel CNN. In 2020 International Conferences on Internet of Things (iThings) (pp. 258-263). IEEE.

\bibitem{r21} Zhang, J., Ghanem, B. (2018). ISTA-Net: Interpretable optimization-inspired deep network for image compressive sensing. In Proceedings of the IEEE conference on computer vision and pattern recognition (pp. 1828-1837).

\bibitem{r32} Zhang, K., Zuo, W., Chen, Y., Meng, D., Zhang, L. (2017). Beyond a gaussian denoiser: Residual learning of deep cnn for image denoising. IEEE transactions on image processing, 26(7), 3142-3155.

\bibitem{r12} Zhang, L. (2015, April). Image adaptive reconstruction based on compressive sensing via CoSaMP. In 2015 2nd International Conference on Information Science and Control Engineering (pp. 760-763). IEEE.

\bibitem{r30} Zhang, Y., Li, K., Li, K., Wang, L., Zhong, B., Fu, Y. (2018). Image super-resolution using very deep residual channel attention networks. In Proceedings of the European conference on computer vision (ECCV) (pp. 286-301).


\end{thebibliography}
\end{document}